\documentstyle[aps,prl,twocolumn,psfig,epsf]{revtex}

\begin{document}
\draft
\twocolumn[
\hsize\textwidth\columnwidth\hsize\csname @twocolumnfalse\endcsname
\title{Phase Transition in a Noise Reduction Model:
Shrinking or Percolation?}
\author{J. van Mourik$^{1,3}$, K. Y. Michael Wong$^2$ and D. Boll\'e$^1$}
\address{$^1$ Instituut voor Theoretische Fysica, 
Katholieke Universiteit Leuven, B-3001 Leuven, Belgium.}
\address{$^2$ Department of Physics, The Hong Kong University of Science
and Technology, Clear Water Bay, Kowloon, Hong Kong.}
\address{$^3$ Istituto Nazionale di Fisica della Materia (INFM), Trieste, 
Italy.}

\maketitle

\begin{abstract}
A model of noise reduction (NR) for signal processing is introduced. 
Each noise source puts a symmetric constraint on the space of the signal 
vector within a tolerable overlap. When the number of noise sources 
increases, sequences of transitions take place, causing the solution space 
to vanish. 
We found that the transition from an extended solution space 
to a shrunk space is retarded because of the symmetry of the constraints, 
in contrast to the analogous problem of pattern storage. 
For low tolerance, the solution space vanishes by volume reduction, 
whereas for high tolerance, the vanishing becomes more and more like 
percolation. The model is studied in the replica symmetric, 
first step and full replica symmetry breaking schemes.
\end{abstract}
\vskip 3mm
]

In the past few years, the statistical mechanics of disordered systems
has been frequently applied to understand the macroscopic behavior
of many technologically useful problems, such as optimization
(e.g. graph partitioning and traveling salesman) \cite{mpv},
learning in neural networks \cite{hkp},
error correcting codes \cite{ecode}
and the $K$-satisfiability problem \cite{mz}.
One important phenomenon studied by this approach
is the phase transitions in such systems, 
e.g. the glassy transition in optimization when the noise temperature
of the simulated annealing process is reduced,
the storage capacity in neural networks
and the entropic transition in the $K$-satisfiability problem.
Understanding these transitions are relevant to the design
and algorithmic issues in their applications.
In turn, since their behavior may be distinct from conventional
disordered systems, the perspectives of statistical mechanics are widened.

In this paper we consider the phase transitions
in noise reduction (NR) techniques in signal processing.
They have been used in a number of applications
such as adaptive noise cancelation, echo cancelation,
adaptive beamforming and more recently, blind separation of signals
\cite{haykin,blind}.
While the formulation of the problem depends on the context,
the following model is typical of the general problem.
There are $N$ detectors picking up signals mixed with
noises from $p$ noise sources.
The input from detector $j$ is $x_j\!=\!a_jS\!+\!\sum_\mu\xi^\mu_j n_\mu$,
where $S$ is the signal,
$n_\mu$ for $\mu\!=\!1,..,p$ is the noise from the $\mu$th noise source,
and $n_\mu\!\ll\!S$.
$a_j$ and $\xi^\mu_j$ are the contributions of the signal and the $\mu$th 
noise source to detector $j$.
NR involves finding a linear combination of the inputs so that the noises 
are minimized while the signal is kept detectable. Thus, we search for an 
$N$ dimensional vector $J_j$ such that the quantities $\sum_j 
\xi^\mu_j J_j$ are minimized,
while $\sum_j a_j J_j$ remains a nonzero constant. 
To consider solutions with comparable power, 
we add the constraint $\sum_j J_j^2 = N$. 
While there exist adaptive algorithms for this objective \cite{haykin}, 
here we are interested in whether the noise can be intrinsically kept
below a tolerance level after the steady state is reached,
provided that a converging algorithm is available.

When both $p$ and $N$ are large, we use a formulation with 
normalized parameters. Let $h^\mu$ be the local fields for the
$\mu$th source defined by 
$h^\mu \equiv \sum_j \xi^\mu_j J_j/\sqrt N$.
Learning involves finding a vector $J_j$ such that the following 
conditions are fulfilled. 
(a) $|h^\mu|<k$ for all $\mu\!=\!1,..,p$, where $k$ is the tolerance bound. 
We assume that the vectors $\xi_j^\mu$ are randomly distributed, with 
$\left\langle\!\left\langle\xi^\mu_j\right\rangle\!\right\rangle\!=\!0$, 
and 
$\left\langle\!\left\langle\xi^\mu_i\xi^\nu_j\right\rangle\!\right\rangle
\!=\!\delta_{ij}\delta_{\mu\nu}$. 
Hence, they introduce symmetric constraints to the solution space. 
(b) The normalization condition $\sum_j J_j^2\!=\!N$. 
(c) $|\sum_j a_j J_j/\sqrt N|\!=\!1$; 
however, this condition is easily satisfied: 
if there exists a solution satisfying (a) and (b) 
but yields $|\sum_j a_j J_j/\sqrt N|$ different from 1, 
it is possible to make an adjustment of each component $J_j$ 
proportional to ${\rm sgn}a_j/\sqrt N$. 
Since the noise components $\xi_j^\mu$ are uncorrelated with $a_j$, 
the local fields make a corresponding adjustment of the order $1/\sqrt N$, 
which vanishes in the large $N$ limit. 
The space of the vectors $J_j$ satisfying the constraints (a) and (b) 
is referred to as the {\it version space}.

This formulation of the problem is very similar to that of pattern 
storage in the perceptron with continuous couplings \cite{Ga}. 
However, in the perceptron the constraints (a) are $h^\mu\!>\!k$, while 
there is an extra inversion symmetry in the NR model: the version space is 
invariant under $\vec J\to -\vec J$. 
We can also consider the NR model as a simplified version of the perceptron 
with multi-state output, in which the values of local fields for each 
pattern are bounded in one of the few possible intervals. In the present 
model, all local fields are bounded in the symmetric interval $[-k,k]$.  
This symmetry will lead to very different phase behavior, although it
shares the common feature that the version space is not connected or not 
convex, with other perceptron models, e.g. errors were allowed \cite{ET}, 
couplings were discrete \cite{KM} or pruned \cite{KGE}, transfer functions 
were non-monotonic \cite{BE}. 

When the number of noise sources increases, the version space is reduced 
and undergoes a sequence of phase transitions, causing it to disappear 
eventually. These transitions are observed by monitoring the evolution of
the overlap order parameter $q$, which is the typical overlap between two 
vectors in the version space. For few noise sources, the version space is 
extended and $q=0$. When the number of noise sources $p$ increases, 
the number of constraints increases and the version space shrinks.
 
One possible scenario is that each constraint reduces the volume of the 
version space, and there is a continuous transition to a phase of nonzero 
value of $q$. Alternatively, each constraint introduces a volume reduction 
resembling a percolation process, in which the version space remains 
extended until a sufficient number of constraints have been introduced, 
and the version space is suddenly reduced to a localized cluster. 
This may result in a discontinuous transition from zero to nonzero $q$. 
We expect that the transition takes place when $p$ is of the order $N$, 
and we define $\alpha\equiv p/N$ as the noise population. 
When $\alpha$ increases further, $q$ reaches its maximum value of 1 
at $\alpha = \alpha_c$, which is called the critical population.
The purpose of this paper is to study the nature 
and conditions of occurrence of these transitions.

We consider the entropy ${\cal S}$, which is the logarithm of the volume 
of the version space and is self-averaging. Using the replica method, 
${\cal S}=\lim_{n\to0}(\left\langle \!\!\!\left\langle{\cal V}^n\right
\rangle\!\!\!\right\rangle\!-\!1)/n$, and we have to calculate
$\left\langle \!\!\!\left\langle{\cal V}^n\right\rangle\!\!\!\right\rangle$ 
given by
\begin{equation}
      \left\langle \!\!\!\left\langle\prod_{a=1}^n \int\prod_{j=1}^N dJ^a_j
      \delta(\sum_{j=1}^N J^{a2}_j\!-\! N)\prod_{\mu=1}^p \theta
      (k^2\!-\!{h^\mu_a}^2)\right\rangle\!\!\!\right\rangle,
\label{pr:1}
\end{equation}
with $h^\mu_a\!\equiv\!\sum_j J^a_j\xi^\mu_j/\sqrt N$. 
Averaging over the input patterns, 
and using the Gardner method \cite{Ga}, 
we can rewrite (\ref{pr:1}) as  $\left\langle
\!\!\!\left\langle{\cal V}^n\right\rangle\!\!\!\right\rangle=\int
\prod_{a<b=1}^n dq_{ab}\exp(Ng)$. The overlaps between the coupling vectors 
of distinct replicas $a$ and $b$: $q_{ab}\equiv{\sum_{j=1}^N }J^a_jJ^b_j/N$, 
are determined from the stationarity conditions of $g$.

Due to the inversion symmetry of the constraints, it always has the 
all-zero solution ($q_{ab}\!=\!0,\forall a\!<\!b$), but it becomes locally 
unstable at a noise population
\begin{equation}
	\alpha_{\rm AT}(k)
	={\pi\over2}{{\rm erf}({k\over\sqrt{2}})^2\over k^2
        \exp(-k^2)}\ .
\label{pr:4}
\end{equation}
For $\alpha\!>\!\alpha_{\rm AT}$, the simplest solution assumes  
$q_{ab}\!=\!q\!>\!0$. This replica symmetric solution (RS), 
however, is not stable against replica symmetry breaking (RSB) 
fluctuations for any $q\!>\!0$. Hence, (\ref{pr:4}) is an Almeida-Thouless 
line \cite{mpv}, and RSB solutions in the Parisi scheme \cite{mpv} have 
to be considered.

The transition of $q$ from zero to nonzero is absent in the problem of 
pattern storage in the perceptron, where $q$ increases smoothly from zero 
when the storage level $\alpha$ increases \cite{Ga}. 
Rather, the situation is 
reminiscent of the spin glass transition in the Sherrington-Kirkpatrick (SK)
model, which does possess an inversion symmetry \cite{mpv}.

The phase diagram is discussed in the following 3 schemes:


{\it 1. RS solution ({\rm RS}, superscript $^{(0)}$):} 
It provides a good approximation of the full picture, 
which will be described later. 
The RS solution is given by $q=q_{\rm EA}$, 
where $q_{\rm EA}$ is the Edwards-Anderson order parameter \cite{mpv}. 
Close to the AT line, $q\sim\!t$, 
where $t\!\equiv\!(\alpha\!-\!\alpha_{\rm AT})/\alpha_{\rm AT}\!\ll\!1$.
The critical population, obtained in the limit $q\!\to\!1$, is
\begin{equation}
	\alpha^{(0)}_c(k)\!=\!\!\left(\!(1\!+\! k^2)(1\!-\!{\rm erf}
	({k\over\sqrt{2}}))\!-\!\sqrt{{2\over\pi}}k\exp({-k^2\over2}\!)\!
        \right)^{-1}\!\!\!\!\!\!\!.
\label{pr:5}
\end{equation}
Two features are noted in the phase diagram in Fig. 1:

{\it a) The critical population line crosses the AT line:} 
When $k\!=\!0$, the version space is equivalent to the solution of $p$ 
homogeneous $N$-dimensional linear equations. 
Hence it vanishes at $p\!=\!N$, or $\alpha_c^{(0)}\!=\!1$. 
The small $k$ expansion of (\ref{pr:4}) and (\ref{pr:5}) gives
\begin{equation}
	\alpha_{\rm AT}(k)\simeq1\!+\!2k^2/3\ <\
	\alpha^{(0)}_c(k)\simeq1\!+\!4k/\sqrt{2\pi}\ , 
\label{pr:6}
\end{equation}
and $q$ increases smoothly from 0 at $\alpha_{\rm AT}$ to 1 at 
$\alpha_c^{(0)}$. However, in the large $k$ limit, the critical population 
grows exponentially with the square of the tolerance $k$, 
and the reverse is true:
\begin{equation}
      \alpha_{\rm AT}(k)\simeq{\pi\over2}{\exp(k^2)\over k^2}>
      \alpha^{(0)}_c(k)\simeq\sqrt{\pi\over2}{k^3\over2}\exp({k^2\over2})
      \ .
\label{pr:7}
\end{equation}

{\it b) The first order transition line takes over the AT line:} 
The paradox in {\it a)} is resolved by noting that for a given $k$,
multiple RS solutions of $q$ may coexist for a given $\alpha$. 
Indeed, $\alpha^{(0)}(k,q)$ is monotonic in $q$ for 
$k\!\le\!k^{(0)}_c\!=\!2.89$, and not so otherwise. 
Among the coexisting stable RS solutions for $k > k^{(0)}_c$, the one 
with the lowest entropy is relevant. Hence, there is a first order transition
in $q$ at a value $\alpha^{(0)}_1(k)$, determined by the vanishing of 
the entropy difference between the coexisting stable solutions. 
The jump in $q$ across the transition widens from 0 at $k=k^{(0)}_c$, 
and when $k$ increases above $k^{(0)}_0=3.11$, 
$q$ jumps directly from zero to nonzero at the transition point. 
The line of $\alpha^{(0)}_1(k)$ starts from $k=k^{(0)}_c$; 
it crosses the line of $\alpha_{\rm AT}(k)$ at $k^{(0)}_0$, 
replacing it to become the physically observable phase transition line.
In the limit of large $k$, the first order transition line is given by 
$\alpha^{(0)}_1(k)\!=\!0.94 \alpha^{(0)}_c(k)$, 
where $q$ jumps from 0 to a high value of $1\!-\!0.27k^{-2}$. 

The phase diagram illustrates the nature of the phase 
transitions. For low tolerance $k$, each constraint results in a 
significant reduction in the version space, and there is a continuous 
transition to a phase of nonzero value of $q$. For high tolerance $k$, 
each constraint introduces a volume reduction which is less significant,
resembling a percolation process, in which the transition of $q$ from zero 
to nonzero is discontinuous. 
However, even in the large $k$ limit, the region of high $q$ spans a 
nonvanishing range of $\alpha$ below the critical population, 
and the picture of percolation transition 
has to be refined in the next approximation.

\begin{figure}
\centering
\centerline{\psfig{figure=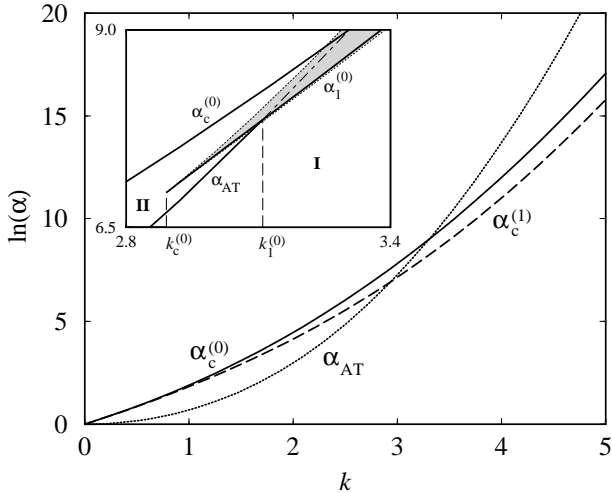,height=7cm}}
\caption{Comparison of $\alpha_c^{(0)}$, 
$\alpha_c^{(1)}$ and $\alpha_{\rm AT}$. 
Inset: the picture within RS near $k_c^{(0)}$ and $k_0^{(0)}$,
with 2 areas: {\bf I)} $q=0$ and {\bf II)} $q>0$.
Shaded area: multiple solutions of $q$.}
\end{figure}

{\it 2. First step RSB approximation ({\rm RSB}$_1$, superscript $^{(1)}$):} 
Here the $n$ replicas are organized into clusters, each with $m$ 
replicas. $q_{ab}\!=\!q_1$ for replicas in the same cluster, and 
$q_{ab}\!=\!q_0$ otherwise, and in the limit $n\to 0$, $\ 0\!<\!m\!<\!1$ 
for analytic continuation. 

Two $(q_1,q_0,m)$ solutions exist just above the AT line:
\newline
($q_{\rm EA},{1\over3}q_{\rm EA},{2\over3}\!m^{(\infty)}$), 
($q_{\rm EA},0,{1\over2}m^{(\infty)}$), 
where (see later) $m^{(\infty)}$ 
is the position of the turning point in the Parisi function of the full 
RSB solution. Only the $(q_0\!>\!0)
$-solution is stable with respect to fluctuations of $q_0$.

The features in the phase diagram in Fig. 2 are:

{\it a) First order transition:} 
For $k\!>\!k^{(1)}_c\!=\!2.31$ multiple solutions exist,
and there is a first order transition of $q$ at $\alpha^{(1)}_1(k)$. 
This line starts from $k=k^{(1)}_c$, and crosses the line of 
$\alpha_{\rm AT}(k)$ at $k^{(1)}_0=2.61$, to become the physically 
observable phase transition line.
For large $k$, $q_1\!=\!1\!-\!1.23/(k\ln k)^2$ at the transition, and 
$\alpha^{(1)}_1(k)\!=\!(1\!-\!0.33/\ln k)\alpha^{(1)}_c(k)$.
Hence, the gap between the lines of $\alpha^{(1)}_1(k)$ and 
$\alpha^{(1)}_c(k)$ vanishes, and the extended phase disappears 
near the critical population via a narrow range of localized phase,
resembling a percolation transition. However, the resemblance is not 
generic, since the approach of the two lines is logarithmically slow.

{\it b) Reduced critical population:} 
The critical population is reduced compared with $\alpha^{(0)}_c(k)$. 
In the large $k$ limit, 
\begin{equation}
\alpha^{(1)}_c(k)\simeq\sqrt{\pi/2}\ k\ \ln k\ \exp(k^2/2).
\label{pr:8}
\end{equation}

{\it c) Transition of $q_0$:} 
The regime of nonzero $q_1$ and $q_0$ spans the region 
of low $k$ below the critical population. 
At larger values of $k$, however,
there is a line $\alpha^{(1)}_m$ where $q_0$ becomes zero. The line 
starts from $k^{(1)}_{c,m}=2.37$ on the line of critical population, 
and ends at $k=k^{(1)}_0=2.61$ on the line of first order transition.
Beyond this line in the localized phase, only the solution with $q_0=0$ 
exists, 
reflecting that constraints with larger $k$ are more likely to reduce the 
size of local clusters, but less likely to change the extended distribution
of the clusters.

The formation of localized clusters at the percolation transition 
is illustrated by the evolution of the overlap distribution. 
For $k>k^{(1)}_0$ at the first order transition line, 
the parameter $m$ decreases smoothly, with increasing $\alpha$, 
from 1 across the phase transition line. This is analogous to the 
RS-RSB$_1$ transition in the random energy model \cite{mpv,fh} and the high
temperature perceptron \cite{gr}. Since the overlap distribution is 
$P(q)\!=\!m\delta(q\!-\!q_0)\!+\!(1\!-\!m)\delta(q\!-\!q_1)$, the 
transition is continuous in the overlap distribution, although 
discontinuous in terms of $q_0$. At the transition, the 
statistical weight of the $q=q_0=0$ component decreases smoothly,
while that of the $q=q_1$ component increases from zero.
When two points in the version space 
are sampled, they have a high probability to share an overlap $q_0$,
meaning that they belong to different clusters. Hence the version space 
consists of many small clusters. Furthermore, since $q_0=0$, the clusters 
are isotropically distributed. In contrast, for the transition at low $k$, 
the overlap has the same form as the SK model, i.e. a high probability 
being $q_1$, implying that it consists of few large clusters.

\begin{figure}
\centering
\centerline{\psfig{figure=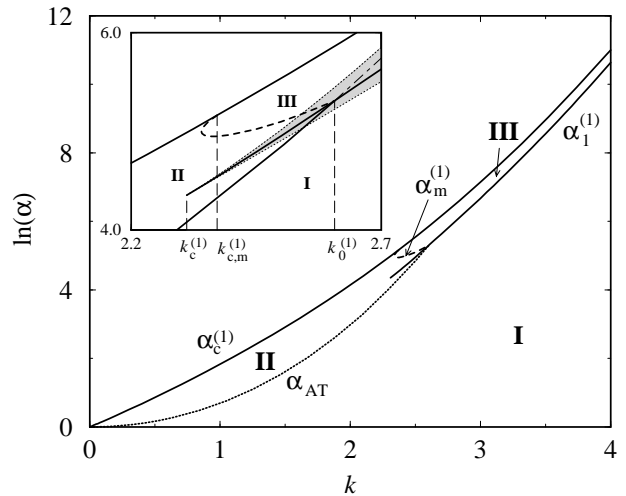,height=7cm}}
\caption{The picture within RSB$_1$, with the three  areas: 
{\bf I)} $q_0=q_1=0$, {\bf II)} $0<q_0<q_1<1$ and {\bf III)} $0<q_1<1$,
$q_0=0$.
Inset: the region near $k_c^{(1)}$ and $k_0^{(1)}$.
Shaded area: multiple solutions of $q_1$ and $q_0$.}
\end{figure}

{\it 3. Infinite step RSB solution 
({\rm RSB}$_\infty$, superscript $^{(\infty)}$):}
This is introduced since the RSB$_1$ solution is found to be unstable against
further breaking fluctuations. 
Here the $n$ replicas are organized into 
hierarchies of clusters. In the $i$th hierarchy, the clusters are of size 
$m_i$, and $q_{ab}\!=\!q_i$. In the limit $n\!\to\!0$, $0\!<\!\cdots\!<\!
m_i\!<\!\cdots\!<\!1$ for analytic continuation. In the Parisi scheme 
\cite{mpv}, the overlap $q_{ab}$ is represented by the Parisi function 
$q(x)$, where $x$ is the cumulative frequency of $q$, or 
$P(q)\!=\!dx(q)/dq$. We have only obtained solutions for 
$\alpha$ just above $\alpha_{\rm AT}$:
\begin{equation}
        q(x)=t\ q_p\min(x/m^{(\infty)},1)\ .
\label{pr:10}
\end{equation}
As shown in Fig. 3, the Parisi function $q(x)$ is very similar to that of 
the SK model without external field near the critical temperature 
\cite{mpv}.

For the behavior of the RSB$_\infty$ solution 
far away from the AT line, we only give some qualitative features of the 
phase behavior. 
Below the AT line we have the extended phase characterized by $q(x)\!=\!0$. 
For small $k$ the AT line lies under the 
$\alpha_c^{(\infty)}(k)$ line, because of (\ref{pr:6}) and 
$\alpha^{(0)}_c\!\!-\!\alpha_c^{(\infty)}\!\simeq\!{\cal O}(k^2)$. 
Hence for $\alpha$ increasing above 
$\alpha_{\rm AT}$, there is a continuous transition of $q(x)$ given by 
(\ref{pr:10}), and $q_{\rm EA}\to1$ for 
$\alpha\!\to\!\alpha_c^{(\infty)}(k)$. 

For large $k$ the AT line runs below the $\alpha_c^{(\infty)}(k)$ line, 
because $\alpha^{(\infty)}_c(k)\!\leq\!\alpha^{(1)}_c(k)\!<\!
\alpha^{(0)}_c(k)$. Hence the $\alpha_c^{(\infty)}(k)$ line must intersect the AT line, 
and there must exist a critical $k^{(\infty)}_c$ above which there is a 
first order transition from a solution with low (or zero) $q_{\rm EA}$ to one 
with high $q_{\rm EA}$ (denoted by $\alpha^{(\infty)}_1(k)$). 
This line intersects the $\alpha_{\rm AT}(k)$ line at a value 
$k^{(\infty)}_0$, replacing it to be the physically observable phase 
transition line. The critical population $\alpha^{(\infty)}_c(k)$ is obtained 
by taking the limit $q_{\rm EA}\to1$. 

We expect that the picture of version space percolation for large $k$ 
continues to be valid in the RSB$_\infty$ ansatz. This means that the 
$\alpha_1^{(\infty)}(k)$ line will become arbitrarily close to the 
$\alpha_c^{(\infty)}(k)$ line.  When $\alpha$ increases above 
$\alpha_1^{(\infty)}(k)$, $q(x)$ will deviate appreciably from the zero 
function only in a narrow range of $x$ near 1. This means that the overlap 
$q$ is zero with a probability almost equal to 1, and nonzero with a 
probability much less than 1. The corresponding picture is that the version 
space consists of hierarchies of localized clusters which are scattered in 
all directions of an $N$ dimensional hypersphere. Again, the transition is 
continuous in terms of the overlap distribution, though not so in terms of 
$q(x)$. 

\newsavebox{\all}
\sbox{\all}
    {\setlength{\unitlength}{0.5mm}
     \begin{picture}(110,55)
        \put(5  ,5 ){\line(0,1){50 }}
        \put(5  ,5 ){\line(1,0){75 }}
        \put(90 ,5 ){\line(1,0){15 }}
        \put(105,5 ){\line(0,1){50 }}
        \multiput(80, 5)(5,0){2}{\line(1,0){3}}
        \put(112,35){\makebox(0,0)[cc]{$q_{EA}$}}
        \put(-2 ,35){\makebox(0,0)[cc]{$q(x)$}}
        \put(5  ,0 ){\makebox(0,0)[cc]{$0$}}
        \put(68 ,0 ){\makebox(0,0)[cc]{$m^{(\infty)}$}}
        \put(105,0 ){\makebox(0,0)[cc]{$1$}}
        \put( 55,-5){\makebox(0,0)[cc]{$x$}}
        \multiput(45,35)(4,0){ 5}{\line(1,0){2}}
        \multiput(45,15)(0,4){ 5}{\line(0,1){2}}
        \multiput(5 ,15)(4,0){10}{\line(1,0){2}}
        \put(48 ,15){\makebox(0,0)[cc]{$_b$}}
        \multiput(35,5 )(0,2){15}{\line(0,1){1}}
        \multiput(35,35)(2,0){ 5}{\line(1,0){1}}
        \put(38 ,8){\makebox(0,0)[cc]{$_c$}}
        \multiput( 7,35)(6,0){5}{\line(1,0){3}}
        \multiput( 5,35)(6,0){5}{\line(1,0){1}}
        \put(15,32){\makebox(0,0)[cc]{$_d$}}     
        \put(65 ,35){\line(1,0){15 }}
        \put(90 ,35){\line(1,0){15 }}
        \put(5  ,5 ){\line(2,1){60 }}
        \multiput(80,35)(5,0){2}{\line(1,0){3}}
        \multiput(65, 5)(0,5){6}{\line(0,1){3}}
        \put(55 ,27){\makebox(0,0)[cc]{$_a$}}
     \end{picture}
    }
\newcommand {\ALL}{\mathord{\!\usebox{\all}}}

\begin{eqnarray}
\begin{array}{l}\ALL\end{array}                                 \nonumber
\end{eqnarray}
\begin{figure}
Fig. 3. The RSB$\infty$ solution near the AT line (a).
For comparison, the RSB$_1$ solution (b: nonzero $q_0$, c: zero $q_0$)
and the RS solution (d) are also plotted.
\end{figure}

The phase transitions observed here have implications to the NR problem. 
In the extended phase for low values of $\alpha$, an adaptive algorithm 
can find a solution easily in any direction of the $N$ dimensional 
parameter space. In the localized phase, the solution can only be found in 
certain directions. If the tolerance $k$ is sufficiently high, there is a 
jump in the overlap distribution, which means that the search direction is 
suddenly restricted on increasing $\alpha$. For very high tolerance $k$, 
the picture of percolation transition applies, so that even though 
localized clusters of solutions are present in all directions, it is 
difficult to move continuously from one cluster to another without 
violating the constraints.

Our results explain the occurrence of RSB in perceptrons with {\em 
multi-state} and {\em analog} outputs, when there are sufficiently 
many intermediate outputs \cite{BvM,BKvM}. In these perceptrons, the 
patterns to be learned, can be separated into two classes: $p_{\rm int}$ 
have zero (intermediate) output, $p_{\rm ext}\!\equiv\!(p\!-\!p_{\rm int})$ 
have positive or negative (extremal) output. The version space formed by 
the intermediate states has the same geometry as the model studied here, 
and consists of disconnected clusters for sufficiently large $p$.

We thank R. K\"uhn for informative discussions, 
M. Bouten, H. Nishimori and P. Ruj\'an for critical comments. 
This work is partially supported by the Research Grant Council of Hong Kong
and by the Research Fund of the K.U.Leuven (grant OT/94/9).


\begin{references}
\vspace{-1cm}
\bibitem{mpv} M. M\'ezard, G. Parisi, and M. Virasoro, {\it Spin Glass Theory 
and Beyond} (World Scientific, Singapore, 1987).
\bibitem{hkp} J. Hertz, A. Krogh and R.G. Palmer, {\it Introduction to
the Theory of Neural Computation}, Addison Wesley, Redwood City, 1991.
\bibitem{ecode} N. Sourlas, {\it Nature} {\bf 339}, 693 (1989).
\bibitem{mz} R. Monasson and R. Zecchina, {\it Phys. Rev. Lett.} {\bf 76},
3881 (1996).
\bibitem{haykin} S. Haykin, {\it Adpative Filter Theory}, second edition,
Prentice-Hall, Englewood Cliffs, NJ (1991).
\bibitem{blind} C. Jutten and J. Herault, {\it Signal Processing} {\bf 24}, 
1 (1991).
\bibitem{Ga}
E. Gardner, {\it J. Phys. A} {\bf 21}, 257 (1988).
\bibitem{ET}
R.Erichsen Jr. and W.K. Theumann, {\it J. Phys. A} {\bf 26}, L61 (1993).
\bibitem{KM}
W. Krauth and M. M\'ezard, {\it J. Phys. France} {\bf 50}, 3057 (1987).
\bibitem{KGE}
R. Garc\'es, P. Kuhlman, and H. Eissfeller, {\it J. Phys. A} {\bf 25}, L1335
(1992).
\bibitem{BE}
D. Boll\'e and R. Erichsen, {\it J. Phys. A} {\bf 29}, 2299 (1996).
\bibitem{fh} K. H. Fischer and J. Hertz, {\it Spin Glasses} (Cambridge University Press, Cambridge, U.K., 1991).
\bibitem{gr} G. Gy\"orgyi and P. Reimann, {\it Phys. Rev. Lett.} {\bf 79}, 
2746 (1997).
\bibitem{BvM}
D. Boll\'e and J. van Mourik, {\it J. Phys. A} {\bf 27}, 1151 (1994).
\bibitem{BKvM}
D. Boll\'e, R. K\"uhn, and J. van Mourik, {\it J. Phys. A} {\bf 26}, 3149
(1993).
\end{references}
\end{document}